\documentclass[preprint,amsmath,amssymb,nofootinbib,aps,]{revtex4-1}

\usepackage{graphicx}
\usepackage{dcolumn}
\usepackage{bm}
\usepackage{epigraph}
\usepackage{tikz}
\usepackage{hyperref}
\newcommand{\ee}[1]{\begin{align}#1\end{align}}

\newenvironment{myepigraph}
  {\par\hfill
   \begin{tabular}{@{}r@{\hspace{2em}}}} 
  {\end{tabular}\par\medskip}

\begin{document}
\title{How We Make Sense of the World: \\ Information, Map-Making, and The Scientific Narrative\footnote{To appear in \it{Map and Territory - Exploring the Foundations of Science, Thought and Reality}, ed. by Shyam Wuppuluri, Newton da Costa, and Francisco Antonio Doria (Springer, Frontiers Collection).}}
\author{Marcelo Gleiser}
\email{mgleiser@dartmouth.edu}
 \altaffiliation[Also at the ]{Department of Physics and Astronomy, Dartmouth College.}
\author{Damian Sowinski}%
 \email{Damian.Sowinski@Dartmouth.edu}
\affiliation{%
 Institute for Cross-Disciplinary Engagement\\
 Dartmouth College\\
 Hanover, NH 03755
}%
\date{\today}

\begin{abstract}
Science is a constructed narrative of the natural world based on information gathering and its subsequent analysis. 
In this essay, we develop a novel approach to the epistemic foundations of the scientific narrative, as based on our experiential interactions with the natural world. 
We first review some of the basic aspects of both Bayesian statistics and Shannon's information theory as applied to the construction of meaningful conceptualization of the natural world. 
This conceptualization is rendered through the maps we construct of the world based on our limited knowledge of reality. 
We propose a path from experience to information and physics based on the notion that information is experience that induces change in an Epistemic Agent (EA): the change may be local and focused to a minor aspect of reality or it may be broad and worldview-changing. 
We illustrate our approach through an analysis of a measure of spatial complexity proposed by one of us called Configuration Entropy (CE), and establish a link between experience at the cognitive level and information content, showing that the CE is a quantitative measure of how much information in spatial-complexity the external world hides from an EA.\\
\begin{myepigraph}
{\it All philosophy is based on two things only: curiosity and poor eyesight.}\\
{\it The trouble is, we want to know more than we can see.}\\[1.5ex]
Bernard le Bovier de Fontenelle
\end{myepigraph}

\end{abstract}


\maketitle

\tableofcontents

\section{Introduction}

The physicist Werner Heisenberg, one of the founding fathers of quantum mechanics, put it clearly: {\it What we observe is not Nature, but Nature exposed to our method of questioning.} 
Excluding those lost in solipsistic confabulations, most people would agree that there is a world external to us, a world that we can apprehend, however indirectly and incompletely, through our sensorial experience, augmented (or not) by adequate instruments. 
Science's central goal, in a nutshell, is to construct meaning from what we perceive of the world. 
Given that what we can perceive of the world is contingent on how we look at it, and that the ways we have been looking at the world have changed with technological advances and consequent shifts in perspective, our constructed meaning of the world is a work in progress: ontologies change due to shifts in epistemic strategies. 
What the world is made of and how it operates are notions frequently revised.

Our relationship with the world is affected through the gathering and exchange of information. 
Experiences are events that promote information flow. 
This is true as we trek along a mountain path, as we communicate with another human, as we dream, or as we read gauges in an experiment. 
As information flows, our awareness state gets updated. 
Senses gather outside information which is then processed in our brains, allowing us to evolve our awareness state. 
Thus informed, we make choices: take another step forward along the path; or don't, given that you reached a precipitous cliff. 
From this simple example, we see that the most informative experiences are those that cause sharper changes in our awareness state.

This conceptualization is not limited to humans, being applicable to any agent, animal or machine, capable of awareness, broadly defined here as the ability to sense the environment. 
For example, a thermostat installed in an air conditioner is an awareness-changing device: it gathers information (data) from the outside environment, sensing changes in temperature. 
Such changes, in turn, will affect the air conditioner's awareness state, causing it to adapt its functioning: if the temperature rises, work harder; if it cools, turn off. 
We call such simple agents {\it Map Walkers}. 
They respond to experiences by changing their awareness state but are incapable of self-awareness. 
They dwell on their perceived Map by updating their awareness states through the flow of their experiences. 
However, they cannot conceive a Map. 
Different Map Walkers experience different maps due to their different sensory apparatuses: a bacterium and an earthworm are both Map Walkers, although they move on very different Maps. 
Their experiences of reality are profoundly different.

What differs radically between a human and an air conditioner or an earthworm is the ability to contextualize experience subjectively.
Humans are aware not only of their environment, but also of their own awareness: they are self-aware. 
A machine or simple animal can process but cannot discern and internalize the notion of surprise, but humans can. 
The most dramatic experiences often surprise us and cause sudden, discontinuous jumps in our awareness state.
It is the awareness that we have of these abrupt changes that quantifies surprise.
The more surprising the experience, the higher the jump. 
These are the experiences that carry the most information. 
Map Walkers will, of course, also respond to a sudden change in conditions if they are programmed -- algorithmically or biologically -- to do so. 
Case in point, a self-driving car must react quite quickly if it senses a child running after a ball in front of it. 
A bacterium driven by chemotaxis will change course if it senses a change in nutrient concentration. 
However, we cannot associate the notion of surprise to a pre-programmed response in a Map Walker because they are not aware of the changes in their awareness.
Perhaps the essential difference between a self-aware and an aware agent is that a preprogrammed Map Walker cannot choose what response it will have for a given stimulus beyond some kind of optimization code. 
We can.\footnote{We are thus making a distinction between awareness and self-awareness. For our purposes, awareness is the property of connecting with both the external and internal environment (awareness states) through an exchange of information; self-awareness is the property of having subjective knowledge of one?s own existence. Given the current lack of knowledge on the nature of consciousness, we consider it here as the black box that endows us (and other presumably self-conscious animals or alien creatures) with this subjective capacity.}

In an effort to conceptualize the role of information in our sense-making of the world, we will therefore focus here mostly on self-aware agents. 
We call these complex agents {\it Map Makers}. 
Note that Map Makers are also Map Walkers but clearly not vice-versa.\\

\noindent
{\bf Note to Non-Mathematical Readers:} Sections IV and VI have some formulas, which we use to make our general statements more quantitatively precise. 
If you are not versed in math, don't worry. 
Skip the formulas. 
We tried hard to be as explanatory as possible in the text before and after the equations so that you can still capture the general sense of our ideas.

\section{Beyond Shannon: Bringing On the Subjective}

If information can be found in the change that is induced in an agent's ability to contextualize experiential data, then, at its heart, information is an epistemic, not ontic, quality: Information is not found in the world; it is found in interactions with the world. 
This is in contrast to the way in which information was introduced in Claude Shannon's {\it A Mathematical Theory of Communication}, the paper that gave birth to the field of information theory \cite{Shannon:1948}. 
It is important to recall that Shannon was interested, as his paper bluntly puts it, in communication: data sent from a source to a receiver over some channel. 
To Shannon, information is the reliable transmission of messages over that channel. 
A channel can relay messages perfectly at a rate no greater than its channel capacity, given the optimal coding scheme. 
Here we glimpse a thread that connects Shannon's view of information as being a reliable transmission of messages, and the epistemically subjective notion of information which we began to develop above:
Shannon's result relies on the existence of an internal mechanism, an encoding. 

Here's a useful analogy. 
Imagine someone pouring one liter of water into a funnel. 
The sender is the agent pouring the water, the one liter of water is the message, the funnel is the channel, and the receiver is on the other side of the funnel. 
If water is poured too quickly, it will overwhelm the funnel and leak from the top: the channel is not being used to its maximum capacity and some of the message will be lost. 
If water is poured too slowly, the operation will take longer than needed. 
To maximize the channel's capacity, the sender must pour water into the funnel at the same rate that it can flow from its narrow bottom. 
This is the optimal encoding for this operation and it depends on the sender's prior knowledge of the channel he is operating with. 
Only with that knowledge can the sender optimize his encoding.

Message transmission requires an alphabet. 
What Shannon discovered was that even in a noisy channel there is an optimal -- {\it error-free} -- transmission rate known as the channel capacity. 
An error-free transmission rate is referred to as having perfect fidelity. 
In the analogy above, not a drop of water is lost. 
So, to maximize the efficiency of a transmission, we need to know the alphabet and, in particular, the frequencies that different letters appear. 
This way, if a letter appears more frequently, it will be more efficient to encode it with fewer bits than one that appears only rarely. 
Think of a bit as the fundamental unit of storage, taking on only two possible values: on or off.
In the English language, the letter {\it e} appears about once every 8 letters, while {\it q} appears once every 1000 letters \cite{AtomsAndVoid:English}. 
It would be very wasteful to encode both with the same number of bits. 
Shannon's result then tells us that if we use the natural frequencies of letters in a language with an alphabet $\mathcal A$ with $N$ symbols, $\mathcal A = \{a_1,a_2,\cdots,a_N\}$, each appearing with a frequency $p(a_i)=p_i$, then the optimal encoding scheme will use $\log \frac{1}{p_i}$ bits of information to encode the $i^{\text th}$ letter $a_i$ of the alphabet.
His magnum opus then relates the channel capacity to the average number of bits needed to store the entire alphabet, the Shannon Entropy:
\begin{equation}
S[\mathcal A] = -\sum_{i=1}^N p_i\log p_i
\end{equation}
Here, again, we see subjectivity crawling into Shannon's results. 
Information, as measured by the optimal rate of message transmission, relies on an internal knowledge of the frequencies of letters in a specific language.

We thus see that in Shannon's approach to communication, a knowledge of the distribution of letters in a language is necessary. 
This knowledge is encoded in a probability distribution over the alphabet. 
For practical purposes, these distributions can be found by examining a large corpus of literature, and extracting the frequencies of each letter. 
This is a statement concerning {\it contextuality}: Without the context of the language being used, the maximal rate of message transmission is meaningless. 
Only Map Makers are capable of creating and encoding meaningful messages. 
Map Walkers are locked into their programs and do not search for meaning or experience doubt. 
For them, there is just doing, as their state of awareness is updated. 
An air conditioner may adjust its functioning due to an input from its thermostat, but its communication with its thermostat, their message exchange, is locked into a one-dimensional realm. 
A laptop will perform commands as programmed, but will not initiate its own encoding or willfully depart from its program. 
To be aware as defined above is not enough to be able to question an algorithmic command; to doubt and choose one needs to be self-aware.

Let us turn then to the context, namely the probability distribution of letters in the alphabet. 
Here the term probability relies on a frequentist interpretation: probabilities are the frequencies with which letters occur. 
The frequentist interpretation of probability is useful in this context, as it is in games of chance and any other repeatable experiences. 
However, it relies on a sort of unrealistic notion wherein the full corpus of all possible messages is used to define the probability with which letters occur. 
Realistic senders and receivers are unlikely to have this full corpus. 
They need to use probability distributions that are inferred from incomplete data. 
Map Makers are always partially ignorant to what they are mapping. 
If we consider science as a Map of physical reality, it follows that scientific theories will always be incomplete maps, given that the information we collect from physical reality (the territory) is inferred from incomplete data.

Probability distributions are inferred from data. 
This is important, for it is a statement about the epistemic state of the sender and the receiver. 
The connection here goes beyond probabilities of letters, and brings us to the Bayesian interpretation of probability. 
Unlike the frequentist point of view, the Bayesian perspective concerning probabilities is that they are not absolute quantities forced on thinking things by the external world, but epistemic measures of the strength of belief thinking creatures like ourselves have about the world \cite{Jaynes:2003a}. 
This interpretation has great strength, as it opens the possibility of defining the probability of not just repeatable events, but of unique ones.

The connection between statements about the world, whether they concern repeatable or unique events, and epistemic states will be explored in the next section. 
We will use an idealized thinking entity, that we will call Idealized Epistemic Agent (IEA), to be defined below. 
Propositional logic will be developed to understand how an epistemic agent talks about the world \cite{Enderton:2001}\cite{Cox:1961a}. 
Following the work of Cox, probability theory will be derived from an epistemic theory about belief, and information will be shown to emerge from processes in which epistemic agents have experiences that alter their beliefs \cite{Cox:1946a}.

Using IEAs as idealized Map Makers, the central thesis of the beginning of this chapter, that information is contained in experiences that change us, will be put on a solid quantitative foundation. 
We will derive Shannon's information measure on purely epistemic grounds, by considering how information is hidden from epistemic agents by the world and revealed (partially) through experience.

The transmission of messages is then the continuous conversation that a Map Maker has with the world; the latter sends messages, in the form of experiences, to the former. 
The results of Shannon's theory then become statements concerning the optimal rate at which Map Makers (epistemic agents) can extract information from the continuous stream of experience. 
In other words, they quantify how good a map Map Makers can create. 
Unlike the receiver in the theorems, Map Makers do not have a fixed set of beliefs, a probability distribution. 
Their interaction with information will cause their beliefs to change: maps get updated.

\section{Epistemic Agents as Idealized Map Makers}
Given the obvious difficulties in defining human consciousness, in order to make conceptual progress on how we relate to the world we need to introduce an idealized Map Maker: an Idealized Epistemic Agent (IEA). 
Humans will be imperfect approximations to IEAs in ways that will be clarified in the following. 
Humans would be EAs, not IEAs. 
Jaynes referred to these inferential automatons as robots \cite{Jaynes:2003a}, while Caticha as idealized rational agents \cite{Caticha:2009a}. 
We prefer the term IEA, since it places emphasis on two aspects that are central to the following: thought and agency. 
The first of these is encapsulated by an EA having a web of consistent beliefs \cite{Caticha:2008}. 
The latter is found in an EA's ability to go beyond simply having experiences as do Map Walkers: EAs generate experiments that evolve their beliefs.

Idealized Epistemic Agents are entities that use an idealized, perfect, language to talk about and conceptualize the world around them. 
Unlike human beings, IEAs are infallible in the use of language. 
Their understanding of language is completely structural, and when they analyze the truth or falsity of a statement, they do so in the context of all possible statements that exist in the language. 
When a particular statement's truth value is established by an IEA, all statements that are logically connected to that statement instantly feel that update. 
An IEA cannot hold within itself contradictions, because all possible statements that could be said in the language are always present within the {\it mind} of the IEA. Humans are not so lucky; we do not have the cognitive resources that an IEA has. 
One may consider paradoxical statements, such as {\it this statement is a lie}.
For now we must assume that these types of self referential statements do not exist within the language. 

An IEA's ability to have access to the complete set of statements has profound consequences to the way they make inferences during experiences. 
For us, an experience does not change all our beliefs in an instantaneous fashion, but for an IEA this is not the case. 
Learning that it is raining has no effect on our belief that the Yen is the currency used in Japan. 
An IEA, however, does not see these statements as being completely independent. 
By pinning down the truth value of rain occurring at this very moment, the IEAs beliefs instantly respond, no matter how tenuously connected they may seem to us: 
\begin{center}
\begin{tikzpicture}
\matrix [column  sep=15mm, row sep=3mm, every node/.style={
    shape=rectangle,
    text width=3.75cm,
    minimum height=0.5cm,
    text centered,
    font=\sffamily\small,
    very thick,
    color=black,
    draw=black,
    fill=white,
}] {
  \node (a) {\it It is raining because an abnormal wind system has caused a storm}; \\
  \node (b) {\it This abnormal wind system is the result of a volcano erupting }; \\
  \node (c) {\it Fujiyama has erupted }; \\
  \node (d) {\it There is major tectonic activity occurring along the Ring of Fire}; \\
  \node (e) {\it Massive earthquakes have caused most of Honshu to submerge}; \\
  \node (f) {\it Japan no longer exists as a sovereign state};\\
};

\begin{scope}[->, very thick, black]
  \draw (a) -- (b);
  \draw (b) -- (c);
  \draw (c) -- (d);
  \draw (d) -- (e);
  \draw (e) -- (f);
\end{scope}
\end{tikzpicture}
\end{center}
(With apologies to our Japanese friends!)
This chain is so implausible that the average human being wouldn't register differences in belief in the links beyond the first. 
The effect of an experience to an IEA is a change, however seemingly insignificant, to its belief in everything else. 
It doesn't matter how far-fetched a line of reasoning goes, all that matters to an IEA is the structure of language: An IEA responds to experience through an instantaneous updating of all of its belief. 

Humans require time to process experience and to fully understand the implications it may have on our beliefs. 
Because processing takes time, experiences for us overlap with one another, blurring the borders between moments. Data from multiple experiences might get mixed up, leading to mistakes in our inference schemes. 
This is not the world that IEAs experience. 
Their ability to instantly process ensures that any experiences separated by finite time intervals, however small, will update the complete set of beliefs the IEAs have in the order in which they happen. 
It is important to keep this distinction in mind in what follows.

\section{From Language to Belief}

Language is that which is used to convey states of the world, both to ourselves, and to our IEAs. 
A language is thus a tool to allows an EA to describe not just the state of the world, but all possible states of the world, and therefore all possible worlds. 
A particular state in the world relies on pinning down the truthfulness of certain statements in the language, which is accomplished via a {\it truth assignment}. 
A language must also have the capacity to bring together statements to construct new statements, which is accomplished via {\bf logical connectives}. 
A language is {\it rational} over the possible worlds on which its logical connectives are consistent. (For technical aspects of logical language construction consult \cite{Sowinski:Thesis}.) 
Possible worlds in which this does not hold are {\it irrational} with respect to the language: the language is incapable of consistently describing those worlds. 
Hence even though the machinery of a language can be used to talk about all possible worlds, only those worlds on which the language can say things clearly can truly be spoken of. 

One mustn't confuse the use of the term language in what follows as a case of a natural language such as English, which will be used in examples.
The language that IEAs use is, like them, an idealization. 
It is assumed that this language is sufficient at describing all the experiences that an IEA can have. 
Clearly, that's not the case for any human dialect.

Now, some definitions. 
A language is made of two kinds of statements, {\bf atomic} and {\bf compound}. 
As the name suggests, atomic statements are irreducible, that is, cannot be decomposed into simpler statements: {\it it is raining}, {\it the cat is alive}, and so forth. 
A compound statement could be, {\it it is raining AND the cat is alive.} 
A language thus consists of atomic statements and logical connectives. 
An epistemic state is a truth assignment on the set of atomic statements. 
The set of all epistemic states with all possible truth assignments is called the epistemic realm. 
(In the example above, the epistemic realm would include:  {\it it is raining}; {\it it is not raining}; {\it the cat is alive};{\it the cat is not alive};{\it it is raining AND the cat is alive};{\it it is raining AND the cat is not alive}; etc.. It would not include {\it it is not raining AND cat}.) 
Given its experience of the world, an IEA would assign a truth value to each of these. 
The subset of the epistemic realm on which the language is consistent with its truth assignment is the realm of discourse. 
A language is rational on its realm of discourse.

Compound statements form the bulk of the statements that are used in common discussions between humans. 
If humans disagree on the truth value of even one statement, then they are not talking about the same world. 
Arguments and experiences may cause us to change truth values.  
We need to see how to incorporate these dynamics in IEAs. 
Furthermore, truth values of statements alone are not enough to capture the possibility that we may be ignorant of the truth or falsehood of a statement: I am not sure whether the statement {\it It will rain tomorrow} is true. 
These considerations lead us to another layer of epistemology, where we must leave behind absolutes, and consider possibilities.

Upon fixing a language, a realm of discourse is established concerning what an epistemic agent may now think about. 
There are many possible worlds in the realm of discourse. 
Many such worlds overlap for the same truth value of a particular statement, or set of statements. 
An epistemic agent prioritizes the value of the truth of statements via a belief function. 
More precisely, for every statement $s$, an IEA has a belief about the statement, $b(s|K)$.
Here we have introduced the prior knowledge of the IEA through the variable $K$.
The term $b(s|K)$ should be read as {\it the strength of belief than an epistemic agent with prior knowledge, $K$, has about the statement $s$.}
In this sense, the epistemic agent prioritizes worlds through belief, and considers the ontological state of the world to coincide with the epistemic world which maximizes their belief. 
What an epistemic agent believes the world to be is that which it is. 
This is true for IEAs and for human epistemic agents.
Since beliefs are meant to prioritize statements, they are an ordering imposed by the IEA. 

Beliefs are transitive, in the sense that if the epistemic agent believes $A$ more than $B$, and $B$ more than $C$, then they must believe in $A$ more than $C$. 
This structure restricts the possible objects that one can use to describe beliefs, which is very useful.
The transitive property means that we can represent beliefs as real numbers, which we write as $b(s|K)\in\mathbb R$.
Belief, then, can be represented mathematically as a function from the set of all statements to the real numbers, $\mathbb R$, given the background knowledge of the IEA. 
Ab initio this background knowledge pertains to the choice of logical connectives chosen by the IEA. 
As the IEA incorporates dynamics in the form of experience, background knowledge will grow to include these experiences. 

Though one may at first think of belief as being binary -- one either does or does not believe in a statement $s$ -- this is not the case. 
Beliefs are graded, and the binary nature of belief only comes about due to a self-imposed threshold. 
An IEA says they believe in a statement when their belief function is beyond this threshold. 
Thus, an IEA's belief in a statement could start off at any value. 
Experience may nudge this value by tiny amounts until finally the threshold is reached, and even then it is possible for the IEA to continue to strengthen their belief in a statement, all the while declaring that they believe. 
We therefore postulate that the belief function is a continuous function, the first of the Cox axioms \cite{Cox:1946a}.

What about the range of the belief function?
Can an EA have a belief of $7$ in one statement, and $-\pi$ in another? 
Is there a maximum strenght of belief, or a minimum?
To make things worse, it is unclear whether higher numbers corespond to stronger or weaker belief.
In fact, every EA can have a different range of belief. 
All that matters to an IEA is the ordering of beliefs. 
This freedom in choosing a range needs to be remedied, since in the end we will want IEAs to be able to compare their beliefs; science is, after all, a dialogue not only between the scientist and reality, but between scientist and scientist.
Changing the scale of belief is accomplished via regraduation. 
This is very similar to choosing a different temperature scale in thermodynamics. 
One chooses the scale in order to make things as simple as possible; scales are chosen on pragmatic grounds, as 
when you go to the doctor and there's a form to fill up that says {\it On a scale from $0$ to $10$ how's your level of pain today?} 
The scale restricts everyone to the same range. 
This is what regraduation does, it rescales the beliefs of all IEAs, while maintaining the ordering of belief, so as to make different IEAs share the same scale.
Regraduation is intimately related with the logical connectives of the language, in that the freedom to choose a scale is constrained only by the demand that the belief function works coherently with the structure of the language.
For example, consider a statement and its negation, $\{s,\neg s\}$.
An IEAs strength of belief in one of these depends on its strength of belief in the other.
We can write this relationship as $b(\neg s|K)=F(b(s|K))$, where $F$ is shorthand for the relationship.
Now the structure of the language kicks in.
Since double negation cancels out (think {\it it is not not raining}), we have that $b(s|K) = b(\neg\neg s|K) = F(b(\neg s|K))=F(F(b(s|K))$.
This puts a severe constraint on the form of the relationship $F$, namely that:
\begin{equation}
\label{Babbage's equation}
F(F(x)) = x
\end{equation}
This {\it functional equation} is quite famous, and is known as {\it Babbage's equation}.
Babbage's work on the analytical engine laid the foundation for modern day computers, and the polymath worked on solving this equation as far back as 1815 \cite{Dubbey:1978}. 
Another such functional equation for the relationship $G(x,y)$, which is constrained by the logical connective {\it and}, reads:
\begin{equation}
\label{Associativity equation}
G(G(x,y),z) = G(x,G(y,z))
\end{equation}
This relationship is called the {\it Associativity equation}, and its beautiful solution can be found in Acz{\'e}l's comprehensive lectures on functional equations \cite{Aczel:1966}.

Using the results from the analysis of both (\ref{Babbage's equation}) and (\ref{Associativity equation}), it can be shown that belief in a truthhood must take on the value $1$\cite{Cox:1946a}\cite{Jaynes:2003a}.
Surprisingly, falsehood can either be represented as $0$ or infinity.
Since truth and falsity are typically represented with a $1$ and $0$, respectively, we make the choice of representing the belief in a falsehood (the least amount of belief that one can have in a statement) with a $0$.
This regraduation also creates a particularly recognizable relationship between beliefs constrained by logical connectives, namely that the sum of the belief in a statement and its negation must equal $1$:
\begin{align}
b(s|K)+b(\neg s|K)=1.
\end{align}
A similar result is familiar from the theory of probability.
In fact, the structural relationships between beliefs turn out to be identical to those of probabilities after regraduation, giving weight to the Bayesian interpretation of probability.
Probabilities are not ontic aspects of reality, but epistemic strengths of belief within IEAs.

\section{From Belief to Information}
Our working premise is that information is that which changes belief \cite{Caticha:2009a}. 
For an experience to be informative, the EA having the experience must be changed upon its conclusion: her awareness state gets updated. 
Let us take a look again at the process of having an experience.  
Consider an EA with a prior set of beliefs about different statements $s$ based on her background knowledge $K$, $b(s|K)$. 
After the experience, she has a posterior set of beliefs where the experience $e$ has been incorporated into her background knowledge, $b(s|e\wedge K)$.
The symbol $\wedge$ represents the now expanded background knowledge incorporating the experience $e$ to the previous background $K$. 
This represents an update of the EA's state of awareness. 
The two states (before $e$ and after $e$) are related by the so-called likelihood, $\mathcal L$:
\begin{equation}
b(s|e\wedge K) = \mathcal L(e;s, K)b(s|K)
\end{equation}

Note that the likelihood is not a belief, since it can be less than, greater than, or equal to $1$. 
It is, however, a very important quantity epistemically. 
If, for a given statement $s$, the likelihood is greater than one, then it strengthens the EAs belief in that statement. 
If it is less than one it weakens it. 
It acts on the old belief distribution multiplicatively to create the new belief distribution.

This procedure can be generalized to an arbitrary number of sequential experiences.
These experiences will update the IEAs beliefs. 
One can either consider the experiences sequentially, or treat them all simultaneously as a single experience (for Idealized EAs only, not for human EAs). 
The equivalence of these two is a statement of the {\it holistic nature of experience}: an IEA may partition her experiences in any way she chooses, but this does not affect her final belief. 

The holistic nature of experience assumes an equivalence between a causal (sequential in time) series of experiences and an acausal --{\it all-together}-- simultaneous piling up of experiences. 
Clearly, this points to what is missing in the theory, since if we are to assume that the beliefs of an EA require a physical substrate, then there must be some ontological cost to changing belief. 
Moving through a series of in-between beliefs will then not necessarily have the same cost as moving from the initial to final belief. 
For example, an EA which throws a die will experience a change in belief due to the outcome. 
They may then draw a card from a deck, which, too, will change their belief function. 
For the holistic nature of experience to hold, the resulting belief after both experiences must be the same in both sequential and simultaneous updating. 
In the case where both are performed sequentially, we would expect that the ontological substrate would have undergone more changes to get to its final state than if they were performed simultaneously. 
The excess change in sequential versus simultaneous updating would be the source for this ontological cost difference, which may not be negligible. 
Indeed, any updating operation involves a thermodynamic cost with a resulting entropy increase. 
The holistic nature of experience, as an idealization, assumes the same heat loss in both cases. 
Despite this shortcoming, it will provide us with the conceptual basis to relate Shannon's entropic formula to the information hidden in the world as experienced by an IEA.
The key comes from our working premise: Information is that which changes belief.
We have seen that experience changes belief from prior to posterior via the likelihood function. 
It is to the likelihood function, then, that we must turn our gaze if it is information that we are trying to understand.

\section{Information Hidden in the World}
Assuming the holistic nature of experience, we can also consider how independent experiences affect the likelihood. 
Independent events are ones whose outcomes do not affect each other. 
For example, throwing a die and drawing a card are independent actions. 
The order of the events is irrelevant. 
Picking a dessert and eating dinner can be dependent experiences, since their order influences one another. 
For independent experiences, the likelihood is the same whatever their order. 
Within this framework, we posit that the information contained in an experience $e$, about a statement $s$, to an IEA with background knowledge $K$, is a function $f$ of the likelihood,
\begin{equation}
I_s(e|K)=f[\mathcal L(e;s,K)].
\end{equation}
We can quickly say a few things from this general statement. 
First, since an uninformative experience doesn't change belief, it must contain no information. 
Mathematically: $f(1) = 0$.
Second, as the likelihood changes by an infinitesimal amount, we do not expect the amount of information contained in the experience to change discontinuously, so the function $f$ must be continuous. 
Lastly, the information gathered from independent events must reflect the commutativity of their temporal ordering: $f(xy) = f(x) + f(y)$. 
It can be shown that the only function that satisfies these properties is the logarithm, allowing us to write
\begin{equation}
I_s(e|K) = A \log [\mathcal L (e;s,K)],
\end{equation}
where $A$ is an arbitrary constant.
 
Since the likelihood is the ratio of prior and posterior beliefs, we can use the properties of logarithms to rewrite the information as:
\begin{equation}
\label{change in information}
I_s(e|K) = A \log b(s|e\wedge K) - A \log b(s|K),
\end{equation}
expressing the information gained by the IEA due to experience $e$ as the change between final and initial states motivates the definition of {\bf hidden information}:
\begin{equation}
h(s|K) = -A \log b(s|K).
\end{equation}
Why call this hidden information. Equation \ref{change in information} tells us that information is the negative change in $h$.
So if the information contained in an experience is positive, then $h$ must have decreased; similarly, if the information contained in the experience is negative, then $h$ must have increased.
This is simply a relationship between revealed and hidden information. 
If that which is revealed increases, then that which was hidden has decreased, and vice-versa.
Furthermore, since we would like the total information hidden by the world to be positive, this means that $A$ is a positive constant. 
The arbitrariness of the constant can be absorbed into the arbitrariness of the base of the logarithm being used. We therefore write
\begin{align}
h(s|K) &= -\log b(s|K) \nonumber \\
&\Downarrow \nonumber \\
 I_s(e|K) &= - \Delta h \nonumber\\
 &= h(s|K) - h(s|e \wedge K)\nonumber
\end{align}
When an IEA has an experience $e$ about a statement $s$ there is a change in hidden information. 
This holds for EAs as well. {\it The fact that the map can never be completely faithful to the territory is, within this framework, an expression of the fact that the hidden information can never be zero; the world will always hide information from an EA.}

There is a nuance in the above that must be addressed.
An experience changes not only the belief in a statement, but also belief in the statement's negation. 
Consider the pair: {\it It is raining}  and  {\it it is not raining}. 
If you look outside, your experience of the weather will make you update your belief in both of these simultaneously.
The total hidden information should depend on both $h(s|K)$ and $h(\neg s|K)$. 
Our prior is that we don't know if it's raining or not. 
We need to think about how both of these contribute to the total. 
If the EA has a strong belief in $s$ (``I'm almost certain it's raining''), then there is very little hidden information in the world concerning $s$. 
On the other hand, if the EA has a very low belief in $\neg s$, there is a lot of information hidden in the world concerning $\neg s$. 
These statements seem contradictory. 
Is there a little or a lot of hidden information in the world concerning the pair $\{s, \neg s\}$? 

To settle this question, we might consider the sum of hidden informations. 
Writing $b=b(s|K)$ for brevity, 
\begin{align}
h(s|K) + h(\neg s|K) &= -\log b - \log (1-b)\nonumber\\
&= - \log b(1-b).
\end{align}
This function is symmetric around $b=\frac{1}{2}$ and gets larger as we move away from this value, diverging as $b\rightarrow 1$ or $b \rightarrow 0$. 
So, as the EA becomes more certain in either $s$ or $\neg s$, the sum of hidden informations increases indefinitely. 
This is absurd!
We must remedy it to properly take into account the ignorance of our EA.
Apparently the total hidden information in the world is not just the mere sum of hidden informations.
How then is the total measured? 

Up to this point, we have been considering the pair $\{s, \neg s\}$, and want a measure of total hidden information $H$ conditioned on knowledge $K$, which has the property that as the corresponding belief pair becomes more polarized, the measure should get smaller. 
(The more the EA knows about the weather, the more she knows whether it's raining or not. 
Her belief in one of the two options gets strengthened.) 
When the gap between the two beliefs approaches zero, it is clear that the amount of hidden information should maximize: very sensibly, the state with maximal hidden information corresponds to the maximally ignorant epistemic state where the EA has no preference whatsoever in believing in the truth or falsehood of a statement:
{\it I have no clue whether it's raining or not.}
These arguments can be generalized to many mutually exclusive statements, which we label as $s_i$, so that $b_i=b(s_i|K)$ for brevity. 
 
What other properties is $H$ endowed with? 
Since it depends on the epistemic state of the EA, it should depend on the beliefs of the EA. 
We now assume that the total hidden information can be factored into a piece that depends explicitly on prior knowledge and another piece, the entropy, that depends implicitly on prior knowledge. 
The simplest possible candidate for the explicitly dependent function is belief itself. 
We then have
\begin{equation}
H[\{s_i\}|K] = b(K)S[\{s_i\}|K].
\end{equation}
By considering how information is related to questioning, it can be shown \cite{Sowinski:Thesis} that the implicit piece must be an expectation value over statements:
\begin{align}
S[\{s_i\}|K] &= \sum_i b(s_i|K)h(s_i|K)\nonumber\\
&=-\sum_i b_i\log b_i
\end{align}
Formally, this is the same formula that Shannon used to define the entropy of an alphabet.
The differences here are twofold.
First, belief is playing the role of probability. 
This is not surprising since we saw earlier that beliefs are structurally the same as probabilities, allowing us to use the Bayesian interpretation of probability. 
Secondly, statements about the world are playing the role of the alphabet.
The messages of experience are transcribed into language, which then forms the core of what belief is about.
We do not believe in an experience, since an experience is something that just IS.
We believe in statements about the world that those experiences inform us about.

For an epistemic agent to be truly honest with their beliefs, not allowing herself to be held back by anything other than the information contained in experience, the EA must strive to always have beliefs that maximize the hidden information.
An IEA, of course, does this automatically.
For humans, it is not always so easy to rid onseself of preconceived notions that have no experiential support.
Interestingly, by postulating this {\it method of maximal entropy} (MaxENT), we connect the epistemic ideal of intellectual honesty with the Second Law of Thermodynamics.
The connection between entropy in the epistemic sense developed here, and the thermodynamic sense of Gibbs, led to the resolution of an age old problem in statistical mechanics known as {\it Maxwell's Demon} \cite{Parrondo:2015}\cite{Sagawa:2012}.

\section{Making Sense of the World: The Relevance of Scale}
Confronted with sensory data, an EA is continuously updating her beliefs about the world. 
The inferences made are a result of applying the Bayesian formalism on experience to generate posteriors, which are then the input for the next round of experience. 
The flux of belief is a signature of there being information in the experiences; the experiences are relevant to the agent and his state of awareness is updated. 
As experiences cause certain beliefs to polarize, the amount of hidden information decreases, corresponding to an increase in the certainty that the agent has about statements. 

An EA in the world will come to believe that certain things happen more regularly than others and that certain objects will appear more frequently in their day to day than others. 
We are pretty sure that on our daily commute to the office we will not have the pleasure of seeing a Triceratops. 
We are, however, pretty certain that we will come across some cars, maybe a bus, and quite a few people. 
Events that occur on familiar scales tend to leave us with very polarized beliefs concerning them. 
They don't carry a lot of information. 
Events that happen on scales much smaller or larger leave us with a sense of surprise. 
These scales may be spatial or temporal. Indeed, routine events do little to change belief; the rarer the event the more impactful it will be. 
This was quantified in the previous section as we equated hidden information with Shannon's entropy.

Let's finesse this further by considering our daily interactions with objects which we perceive on length scales close to our own. 
Imagine entering a room for the very first time. 
The walls are painted in a slightly off red color, and there is a desk with a computer. 
There are shelves along the walls filled with books, and posters on the walls. 
Next to the corner closest to us we notice a small spider, dangling from a silken thread, its legs moving frantically in ascent. 
We turn and look more closely at the shelves, noticing books about physics and philosophy, and when we focus back to the desk we become interested in papers on it which seem to be covered with scribbles. 
Closer inspection reveals equations, and as we touch the papers, a pencil rolls from underneath them to the edge of the desk, and falls off the side. 
We pick it up, put it back on the desk, and then stand back, panning our gaze over the room as a whole. 
We imagine the academic that must call this room her home, and all the questions we'd like to ask her about the books and the papers. 
We're standing now close to the corner next to the entrance, when a memory tickles us, and we turn back to see the spider. 
It is no longer there. 
We panic just a bit, brushing ourselves in case the spider jumped on our body in the brief moments that we were taking stock of the office. 
Several brushes and a quick inspection leaves us confident that the spider is most likely not on us, but as we look around frantically we cannot find it. 
Realizing we're being foolish, we smile and look back at the shelves to see if any of the books might be of interest. 
We note a few and remember them. 
If we ever see the owner of the office we will have to ask her about those books. 
We turn and go from whence we came.

What can we learn from this story? 
There are a few things that seem obvious, but should be stated explicitly. 
When we first entered the office, we were attentive to the existence of the shelves and the desk, and to the colors of the walls. 
The largest things in the room caught our attention, and created the skeletal structure of the model in our minds of the room we were in. 
Once our beliefs were polarized about the general shape of the room and its coarsest features, we began to pay more attention to smaller things: 
The types of books and papers on the desk; the motion of the spider alerting us to its presence even though it is much smaller than the scales we are currently investigating. 
After taking note of it we turned to the equations which, too, were small relative to the other objects in the room. 
Why do we do things in this order? 
Why do we investigate starting at the coarsest scales first, and then move to finer scales? 
How can instability, as exemplified by the unexpected presence of the spider, derail us from this general mode of investigation? 

We should look at the informational narrative of the above process, given our framework. 
We began with quite a bit of hidden information when we entered the room. 
Experience quickly remedied that. But why? 
For an answer, we turn to the concept of {\it coarse-graining}. 
A coarse-grained distribution has less hidden information in it, in general, than a fine distribution. 
({\it Look, it's a group of 100 people.}) 
Pinpointing a value in such a distribution seems to not decrease the amount of hidden information by much. 
({\it Look, it's one of these people.}) 
However, the act of conditioning a fine-grained distribution based on the information from a coarse-grained experience will significantly decrease the hidden information in the fine distribution.
({\it Look, people are wearing shirts of different colors and are grouped according to color.}) 
Thus, the act of minimizing hidden information by forcing experiences through actions (looking around the room, gazing at what is on the shelves and on the desk), can be accomplished more efficiently by pinning down coarse-grained features first, and using them to throw away irrelevant fine features. 
The philosopher Evan Thompson referred to this relation with the world as autopoiesis, seen as the {\it dynamic co-emergence of an interiority and exteriority} \cite{Thompson:2007}.

Of course, the spider was not a coarse-grained object in the room, so noticing it seems to be at odds with the interpretation that gaining information in the most efficient manner should progress from coarser to finer scales. 
What made something at a finer scale more relevant to us? 
Well, first off it was in motion. 
Since the rest of the office was in a static state, the best way to proceed in reducing hidden information would be the process of conditioning on coarser experiences. 
Things that move capture our attention because of the instability that change introduces to the static world. 
Abrupt change leads to an abrupt update in the state of awareness.

Ignoring the particulars of different objects that are the same size, what is a good measure for the amount of hidden information to an EA which has just been introduced into a room that contains objects, or distributions of objects, at many different scales? 
Correlations between the scales and positions of objects will need to be considered, since sizes and orientations contribute to the overall information that the EAs want to discover about the general shape of the space that they're in. 
Scales at which correlations are the largest will cause the EA's belief about objects at those scales to polarize the fastest. 
Smaller and larger scales relative to this correlation scale will contain much more hidden information. 
In making sense of the world, the unusual is what makes the difference. 

We are thus led to consider a measure that can capture spatial correlations, such that less correlated scales imply more hidden information. 
This quantity is called Configurational Entropy (CE), introduced by Gleiser and Stamatopoulos in 2012, and inspired by Shannon's information entropy \cite{Gleiser:2012}. 
Essentially, the CE measures the spatial correlation of objects at different length scales. 
It can be used very effectively to detect patterns that jump out of the background at a certain distance scale (a certain size). If all we see (the {\it message}) is a noisy mess, the CE will be large: there are no perceived correlations at any scale and hidden information is large. 
If, on the other hand, there are patterns in space at certain distances, the CE will be smaller and so will be hidden information.
\onecolumngrid
\begin{figure*}
\includegraphics[width = \textwidth]{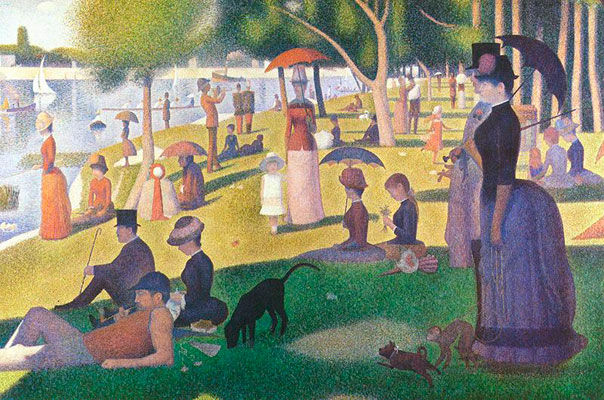}
\caption{Georges Seurat's classic pointillist painting, {\it A Sunday Afternoon on the Island of La Grand Jatte}. Helen Birch Bartlett Memorial Collection, The Art Institue of Chicago.}
\end{figure*}
A possible analogy is George Seurat's pointillist paintings. 
Looked at too closely, all we see are colored points with no discernible pattern. 
Take a step or two back, and patterns begin to emerge until a picture forms in our minds, a scene in a park with people, parasols, trees, and animals. 
Although this is not the proper place to go into the technical details of CE, we can state that it offers a measure of spatial complexity in the physical world based on the concept of hidden information and, thus, on the relation between an EA and the world she perceives through experience. 
In the next section, we sketch the foundational aspects of this relation, which we will present in more detail in a future publication.
For phenomenological applications of CE, look at Refs. \cite{Sowinski:2013}\cite{Sowinski:2015}\cite{Gleiser:2015}\cite{Sowinski:2017}.

\section{Psychophysical Foundations of Configurational Entropy}
For most of human history, there were 1025 stars in the Heavens  \cite{Pratt:Stars}.
These pinpricks of light were all that the human eye could see and write about, and these thousand stars, together with the wanderers ({\it planetos}) and the occasional shooting star, were the sum total of human knowledge about the census of the sky. 
The stars were subdivided into six classes based on how bright they were, with $m=1$ being the brightest, and $m=6$ the faintest. 
It wasn't until the invention of the telescope in 1608 that this catalogue began to increase in size, with dimmer and dimmer stars being discovered. 
Initially, the magnitude scale appeared to be increase linearly with brightness based on the human eye's ability to register light. However, when more sophisticated methods were brought to bare on the field of photometry, it turned out that magnitudes were {\it logarithmically} related to the amount of light being received: stars differing by one magnitude point were twice as bright, those differing by two magnitudes four times, three magnitudes eight times, and so on.

Logarithms are very important in the realm of perception. For two stimuli to register as being different, the senses must perceive them with a {\it just noticeable difference} (JND), a threshold for perception. 
Stimuli that do not create a JND to the senses are perceived as being the same. 
Consider, if you will, two rulers placed before you. One of them is a decimeter in length, while the other a meter. 
Imagine that both are instantaneously increased in length by one centimeter. Perceiving a change in the smaller ruler will be obvious. However,
an EA may or not perceive the change in the larger ruler. 
Had one of the rulers been a kilometer in length, it is certain hat a change by one centimeter would not register in the EA's senses. 

In 1860, Ernst Weber proposed a quantitative relation for the
change in perception, $\Delta \mathcal P$, equating it to both a change in stimulus, $\Delta S$, and the stimulus, $S$ \cite{Fechner:1860}:
\begin{equation}
\Delta \mathcal P \propto \frac{\Delta S}{S}.
\end{equation}
The solution (the change in perception) is a logarithm. 
In that same year, Gustav Fechner tested the relation experimentally. Since then, it's been known as the Weber-Fechner relation (WFR). Equating the experience $e$, with the change in perception $\Delta {\cal P}$, we can write
\begin{equation}
e(S) \propto \log \left (\frac{S}{S_0} \right ),
\end{equation}
where $S_0$ is a baseline stimulus used to calibrate the relationship.

The WFR begins to break down at the boundary of sense perception, though it holds for  each of the senses. 
In particular, we can use it in the context of scale perception as a way to constrain the belief distribution that an EA has about its environment.\footnote{With this relation, we can write that the experience for an EA with prior knowledge $K$ and belief $b(k|K)$ of a given scale (size) $k$ is $e(k) = b(k|K)e(S(k))$, so that the average experience is the sum over all scales $k$, $\langle e\rangle = \sum_k b(k|K)e(S(k))$.}

In the case of spatial perception, we propose that the stimulus is related to the {\it two-point correlation function}, a mathematical quantity that describes how pairs of points are correlated in a spatial environment, peaking at scales typical of most objects or physical properties in that environment. 
For an example of a physical property, consider the temperature at different points in a room. 
If there is little change in temperature from point to point, the two-point correlation will be large, given that most points have similar temperatures and are thus highly-correlated. 
If, instead, the temperature fluctuates randomly from point to point, the two-point correlation will be very small.
More precisely, the two-point correlation function gives the relative power at different scales, which, in our formulation, is related to the strength of the stimulus at different scales. 
(In the example of the temperature in a room, the scales will be related to the different sizes of the volumes in the room that have the same temperature. 
The biggest volumes with the same temperature will have the most power in the two-point correlation function.)

Constraining the hidden information for an IEA by the mean experience of spatial scales using the WFR, one then finds that an IEA should have a belief distribution over spatial scales that is a power law of the power spectrum, coined the modal fraction in Ref. \cite{Gleiser:2012}. 
As shown in that work, the modal fraction is the key ingredient of the Configuration Entropy, introduced in Section VII. Essentially, the modal fraction $f({\bf k})$ gives the relative probability of a given spatial scale over all others. 
If a spatial scale is prominent, it will dominate the modal fraction, which can have a value between 0 and 1.\footnote{Mathematically, the modal fraction is proportional to the {\it power spectrum}, $f({\bf k}) \propto {\cal P}(k)$ and so is directly related to the two-point correlation function. (For the mathematically savvy, the power spectrum is the Fourier Transform of the two-point correlation function.) 
The Configurational Entropy is the hidden information of the MaxEnt distribution under the constraint of average experience $k$, $S_{CE} = - \sum_k f({\bf k})\log [f({\bf k})]$.}
Using this belief distribution, the Configurational Entropy (CE) becomes a quantitative measure of how much information in spatial-complexity the external world hides from an IEA: a world of sameness hides little, while a world rich in spatial patterns at different scales hides a lot. 
In this way, the CE offers a quantitative measure for an IEA's different experiences. 
These experiences are then used by an IEA to construct a map of its perceived physical reality.

\section{Concluding Remarks}

As sentient beings, humans are forever locked within their limited perception of physical reality. 
One may conjecture of an ``ultimate reality,'' what we could call the perfectly complete ontological Territory, but such entity is certainly out of reach, even for Idealized Epistemic Agents. 
At best, we can collect partial information about the Territory as we confront the world with our prior knowledge, working to decrease, through experience and its decoding, the amount of hidden information. 
The Maps we construct are the products of such a process, always works in progress. 
Science is one of such Maps, but certainly not the only one. 
As we discussed, what we sense of reality is, in the parlance of information theory, the message. 
As we experience the world through different stimuli and consequently update our state of awareness, we decrease the amount of hidden information. 
In our framework, information is that which changes belief.

When applied to how an EA senses and moves in the world, the information updating depends on spatial perception. 
We have presented a formalism to describe how this process works by matching a quantity called Configurational Entropy to the hidden information. 
One can think of the Configuration Entropy as a measure of the spatial complexity of an objects and how it relates to other objects nearby. 
As an EA enters a new environment and begins to sense it, she searches for spatial correlations among objects. 
Stimuli that promote the most change in the EA's state of awareness are the ones that carry the most information. 
These tend to be the stimuli that depart most strongly from the average experience. 
According to the Weber-Fechner relation, such change grows with the logarithm of the intensity of the stimulus. 
Using this expression, we were able to show that, under certain assumptions, the hidden information is given by the Configurational Entropy, computed from the spatial correlation between different objects in the room. 
The more varied the room, the greater the stimulus, the richer the experience, and the more hidden information. 
As the EA keeps on exploring and updating her belief, different experiences will decrease the amount of hidden information.

\vspace{0.5 in}
\noindent 
{\bf Acknowledgements}: This work was supported by the Institute for Cross-Disciplinary Engagement at Dartmouth College (http://ice.dartmouth.edu) by a John Templeton Foundation grant.


\appendix
\section{Outline of Derivation of Configurational Entropy}
Consider a stimulus $S$.
This can be anything from sound to light to touch.
Denote an EA's experience of the perception of that stimulus as $e(S)$. 
Now consider changing the stimulus by some small amount, $S\rightarrow S+\delta S$.
The empirical Weber-Fechner relation states that a change in perception is proportional to a change in the stimulus, but inversely proportional to the stimulus itself:
\ee{
\delta e = \eta\frac{\delta S}{S},
}
where $\eta$ is some empirically determined constant of proportionality. 
This implies that the experience is proportional to the logarithm of some power of the stimulus,
\ee{
e(S) = e_0+\log S^\eta.
}
Here $e_0$ is a constant of integration.
It is reasonable that when the stimulus reaches some minimum value, $S_0$, then one can no longer perceive it, thus
\ee{
e(S) = \log\left(\frac{S}{S_0}\right)^\eta.
}

For the experience of spatial scale, the stimulus is proportional to the power spectrum
\ee{
S(k)\propto \mathcal P(k),
}
where $k=\frac{\pi}{L}$ is a wave-mode at some inverse length scale, $L$.
Note that the smallest perceptible scale will correspond to some maximum wave-mode, $k_*$, so that  $S_0\propto \mathcal P(k_*)$.
An EA will have some belief distribution over the importance of scales, $b(k|K)$, based on its background knowledge $K$.
The EA's mean experience of scales will then just be:
\ee{
\langle e\rangle &= \sum_k b(k|K)e(S(k))\nonumber\\
&=\eta \sum_k b(k|K)\log\frac{\mathcal P (k)}{\mathcal P(k_*)}.
}
What distribution should the EA have over scales that is as unbiased as possible considering nothing but mean experience?
To find an answer, we apply the MaxEnt method \cite{Parrondo:2015}.

The total hidden information in spatial scales for the EA can be expressed by the functional:
\ee{
H[\{ b\};\alpha,\beta,\gamma]=-\sum_k b(k)\log b(k)+\alpha \left(1-\sum_k b(k)\right)+\beta\left(\langle e\rangle-\eta \sum_k b(k)\log\frac{\mathcal P (k)}{\mathcal P(k_*)}\right)+\gamma(\cdots),
}
where $\alpha$, $\beta$, and $\gamma$ are Lagrange multipliers that enforce constraints.
The first term is the entropy of the distribution. 
It tries to drive the belief distribution towards uniformity (ignorance).
The second term is the constraint enforcing that the belief distribution be normalized.
The third term is the constraint enforcing mean experience.
The last term is all the constraints that impose the EA's background knowledge, $K$.
Since we are only worried in how mean experience of scale constrains belief, we can set $\gamma$ to zero. 
Varying with respect to the Lagrange multipliers simply reproduces the constraints:
\ee{
\frac{\delta H}{\delta \alpha} = 0&\Rightarrow \sum_kb(k)=1;\\
\frac{\delta H}{\delta \beta} = 0 &\Rightarrow \eta \sum_k b(k)\log\frac{\mathcal P (k)}{\mathcal P(k_*)}=\langle e\rangle.
}
Varying with respect to the belief distribution, we find
\ee{
\frac{\delta H}{\delta b(k)}=-\log{b(k)}-1-\alpha-\beta\eta\log\frac{\mathcal P (k)}{\mathcal P(k_*)}.\nonumber
}
Setting this variation to $0$,
\ee{
b(k) = e^{-(1+\alpha)}\left(\frac{\mathcal P(k)}{\mathcal P(k_*)}\right)^{-\beta\eta}.
}
Imposing the normalization constraint we obtain,
\ee{
b(k) = \frac{\mathcal P(k)^{-\beta \eta}}{\sum_{k'}\mathcal P(k')^{-\beta \eta}}.
}
Choosing $\beta\eta = -1$ \cite{Sowinski:2017b}, we obtain a special case for which the belief distribution is the modal fraction, $f(k)$:
\ee{
b(k) = \frac{\mathcal P(k)}{\sum_k'\mathcal P(k')} = f(k).
}
Plugging this, and the constraints, into the hidden information functional gives us the expression for the Configurational Entropy in terms of the modal fraction from Ref. \cite{Gleiser:2012}:
\ee{
S_C = H[\{b\};\cdots]=-\sum_k f(k)\log f(k). 
}
This results establishes a quantitative link between the psychophysical foundations of spatial perception and the configurational entropy as a measure of hidden information for an EA.

\end{document}